\documentclass[letter]{aa} 
\usepackage{natbib}
\bibpunct{(}{)}{;}{a}{}{,} 

\usepackage{graphicx}
\usepackage{txfonts}
\usepackage{amsmath,amssymb}
\usepackage{mathrsfs}	
\usepackage{color}
\usepackage[bookmarks=true]{hyperref}
\hypersetup{
  colorlinks=true,
  pdfborder=2 2 0.,
  linkcolor=red,
  anchorcolor=black,
  citecolor=blue,
  urlcolor=black,
}
\usepackage{breakurl}

\newcommand{\hi}{\ifmmode{\rm HI}\else{H\/{\sc i}}\fi} 
\newcommand{\Msun}{M_\odot}
\newcommand{\Mstar}{M_\ast}
\newcommand{\jstar}{j_\ast}
\newcommand{\Mh}{M_{\rm h}}
\newcommand{\jh}{j_{\rm h}}
\newcommand{\Rd}{R_{\rm d}}
\newcommand{\de}{{\rm d}}
\newcommand{\Vc}{V_{\rm c}}
\newcommand{\Vf}{V_{\rm f}}
\newcommand{\jd}{j_{\rm d}}
\newcommand{\Md}{M_{\rm d}}
\newcommand{\fdisc}{f_{\rm d}}

\newcommand{\fjs}{f_{j,\ast}}
\newcommand{\fms}{f_{M,\ast}}
\newcommand{\fjd}{f_{j,\rm d}}
\newcommand{\fmd}{f_{M,\rm d}}

\defcitealias{RF12}{RF12}
\defcitealias{SPARC}{LMS16}

\titlerunning{Angular momentum-mass law from dwarf to massive spirals}
\authorrunning{L. Posti, F. Fraternali, E. Di Teodoro \& G. Pezzulli}

\begin{document}

\title{The angular momentum-mass relation: a fundamental law from dwarf irregulars to massive spirals} 

   \author{Lorenzo Posti\inst{1}\fnmsep\thanks{posti@astro.rug.nl},
          Filippo Fraternali\inst{1},
          Enrico M. Di Teodoro\inst{2}
          \and
          Gabriele Pezzulli\inst{3}
          }
   \institute{Kapteyn Astronomical Institute, University of Groningen,
   			  P.O. Box 800, 9700 AV Groningen, the Netherlands
         \and
             Research School of Astronomy and Astrophysics - The Australian National
             University, Canberra, ACT, 2611, Australia
         \and 
             Department of Physics, ETH Zurich, Wolfgang-Pauli-Strasse 27, 8093 Zurich, 
             Switzerland
             }
      \date{Received XXX; accepted YYY}
   
   \abstract{
      		In a $\Lambda$CDM Universe, the specific stellar
            angular momentum ($\jstar$) and stellar mass ($\Mstar$) of a galaxy are 
            correlated as a consequence of the scaling existing for dark matter haloes 
            ($\jh\propto\Mh^{2/3}$). 
            The shape of this law is crucial to test galaxy formation models,
            which are currently discrepant especially at the lowest masses, allowing to
            constrain fundamental parameters, e.g. the retained fraction of angular momentum.
            In this study, we accurately determine the empirical $\jstar-\Mstar$ relation 
            (Fall relation) for 92 nearby spiral galaxies (from S0 to Irr) selected from 
            the {\it Spitzer} Photometry and Accurate Rotation Curves (SPARC) sample in 
            the unprecedented mass range $7 \lesssim \log\Mstar/\Msun \lesssim 11.5$. 
            We significantly improve all previous estimates of the Fall relation
            by determining $\jstar$ profiles homogeneously for all galaxies, using
            extended \hi\ rotation curves, and selecting only galaxies for which
            a robust $\jstar$ could be measured (converged $\jstar(<R)$ radial profile).
            We find the relation to be well described by a single, unbroken power-law 
            $\jstar\propto\Mstar^\alpha$ over the entire mass range, with $\alpha=0.55\pm 
            0.02$ and orthogonal intrinsic scatter of $0.17\pm 0.01$ dex. 
            We finally discuss some implications for galaxy formation models of this
            fundamental scaling law and, in particular, the fact that it excludes
            models in which discs of all masses retain the same fraction of the halo
            angular momentum.
             }
   \maketitle

\section{Introduction} \label{sec:intro}

Mass ($M$) and specific angular momentum ($j=J/M$) are two independent and key galaxy 
properties, subject to physical conservation laws, which are correlated in a fundamental 
scaling relation, the $\jstar-\Mstar$ law. This was first introduced by \cite{Fall1983}, as
a basis for a physically-motivated classification of galaxies, and hence we call it the
\emph{Fall relation} hereafter. Empirically, massive spiral galaxies ($\log\Mstar/\Msun\gtrsim
9$) are found to lie on a power-law relation close to $\jstar\propto\Mstar^{2/3}$ 
\citep[][hereafter \citetalias{RF12}]{RF12}.

In a $\Lambda$ Cold Dark Matter ($\Lambda$CDM) Universe, this fundamental relation
highlights the intimate link between galaxies and their host dark matter haloes: in fact,
the specific angular momentum of haloes scales precisely as their mass to the power $2/3$,
as a result of tidal torques \citep{Peebles69,EfstathiouJones1979}. As highlighted by early 
semi-analytic models \citep{Dalcanton+1997,MMW98},
this connection is mediated by two fundamental physical parameters: $\fms\equiv\Mstar/\Mh$, 
the so-called \emph{global star-formation efficiency}, and $\fjs\equiv\jstar/\jh$, the 
so-called \emph{retained fraction of angular momentum}, which encapsulates several processes 
relevant to galaxy formation, including angular momentum losses due to interactions and 
the possibility that the gas which contributes to star formation does not sample uniformly
the global angular momentum distribution. In particular, the observed Fall relation is key 
to constrain $\fjs$ as a function of other galaxy properties \citep{Posti+2017,Shi+2017}.

Several galaxy formation models are now able to correctly predict the amount of angular
momentum in massive spiral galaxies \citep[e.g.][]{Genel+2015,Teklu+2015,Zavala+2016}.
However, these predictions become rather discrepant and uncertain for the lower mass systems
($\log\Mstar/\Msun\lesssim 9$), where some models predict a flattening of the relation
\citep{Obreja+2016,Stevens+2016,Mitchell+2018} while others do not see any change with
respect to the relation for larger spirals \citep{El-Badry+2018}. These discrepancies are
arising also because observational estimates of the $\jstar-\Mstar$ relation over a wide 
galaxy stellar-mass range are lacking.

The aim of the present Letter is to provide the benchmark for the Fall relation from dwarf
to massive spirals in the local Universe. We use spirals of all morphological types spanning
an unprecedented mass range ($7\lesssim\log\Mstar/\Msun\lesssim 11.5$), using accurate 
near-infrared photometry/\hi\ data to trace the stellar mass/galaxy rotation out to several 
effective radii.
Unlike many previous estimates of massive \citep[\citetalias{RF12};][]{OG14,Cortese+2016} and
dwarfs separately \citep{Butler+2017,ChowdhuryChengular2017}, we homogeneously measure
$\jstar$ profiles for all galaxies and determine the relation using only those with a
converged value of the total $\jstar$.

The paper is organised as follows. Sect.~\ref{sec:data} introduces the dataset used.
Sect.~\ref{sec:fall} explains our method and selection criteria and presents our
determination of the Fall relation. In Sect.~\ref{sec:models} we discuss the implications
of our findings for galaxy formation models. We summarize and conclude in 
Sect.~\ref{sec:concl}.

\section{Data} \label{sec:data}
The sample of spiral and irregular galaxies considered in this work comes from the 
{\it Spitzer} Photometry and Accurate Rotation Curves (SPARC) sample \citep[][hereafter
\citetalias{SPARC}]{SPARC}. 
For these 175 nearby galaxies, from S0 to Irr, surface brightness 
profiles at 3.6 $\mu$m, derived from {\it Spitzer Space Telescope} photometry, and
high-quality neutral hydrogen (\hi) rotation curves, derived from interferometric \hi\ data, 
are available.

Near-infrared profiles best trace the stellar mass distribution \citep[e.g.][]{Verheijen2001}, 
as the mass-to-light ratio at 3.6$\mu$m is nearly constant over a broad range of galaxy masses 
and morphologies \citep[e.g.][]{BelldeJong2001,McGaughSchombert2014}. For this work we assume 
the fiducial values used in \cite{Lelli+2017} for stellar population models with a 
\cite{Chabrier2003} initial mass function: $\Upsilon^{[3.6]}_{\rm b}=0.5$ and 
$\Upsilon^{[3.6]}_{\rm d}=0.7$ for the bulge and disc at 3.6 $\mu$m, respectively\footnote{
These values are comparable with those used by \cite{FR13} in their updated calibration
of the Fall relation with respect to \citetalias{RF12}, who used $\Upsilon^K=1$ (in $K$-band)
for both bulge and disc.
}. 
The photometric profiles have also been decomposed in bulge/disc as described
in \citetalias{SPARC}.

In a disc galaxy most stars are on nearly-circular orbits and rotate with velocities close to
the local circular speed. We use the available \hi\ rotation curves \citepalias[from][]{SPARC}
to trace the circular velocity, then we apply a correction for the asymmetric drift \citep[]
[\S 4.8.2]{BT08} to get the stellar rotation curve (see Sect.~\ref{sec:ad}).

\section{The specific angular momentum-mass relation} \label{sec:fall}

If a galaxy is axisymmetric and rotates on cylinders about its symmetry axis, then
the \emph{specific stellar angular momentum} $\jstar\equiv |{\bf J}_\ast|/\Mstar$ within 
the radius $R$ from the galactic centre writes as

\begin{equation} \label{eq:jstar}
\jstar(<R) = \frac{\int_0^R \de R' \,\, R'^2\,\Sigma_\ast(R')\,V_{\rm \ast,rot}(R')}
			{\int_0^R \de R' \,\, R'\,\Sigma_\ast(R')},
\end{equation}

\noindent where $\Sigma_\ast(R) = \Upsilon^{[3.6]}_{\rm b}\,I_{\rm b}(R) + 
\Upsilon^{[3.6]}_{\rm d}\, I_{\rm d}(R)$ is the surface stellar mass density, with $I_{\rm b}$
and $I_{\rm d}$ being the surface brightnesses of the bulge/disc at 3.6 $\mu$m, and $V_{\rm 
\ast,rot}$ the stellar rotation curve. The total specific stellar angular momentum is
$\jstar\equiv\jstar(<R_{\rm max})$, where $R_{\rm max}$ is the outermost radius at which
$\Sigma_\ast$ is measured. We compute the galaxy stellar mass as 
$\Mstar=2\pi\int_0^{R_{\rm max}} \de R'\,R'\Sigma_\ast(R')$.

\begin{figure}
\includegraphics[width=0.5\textwidth]{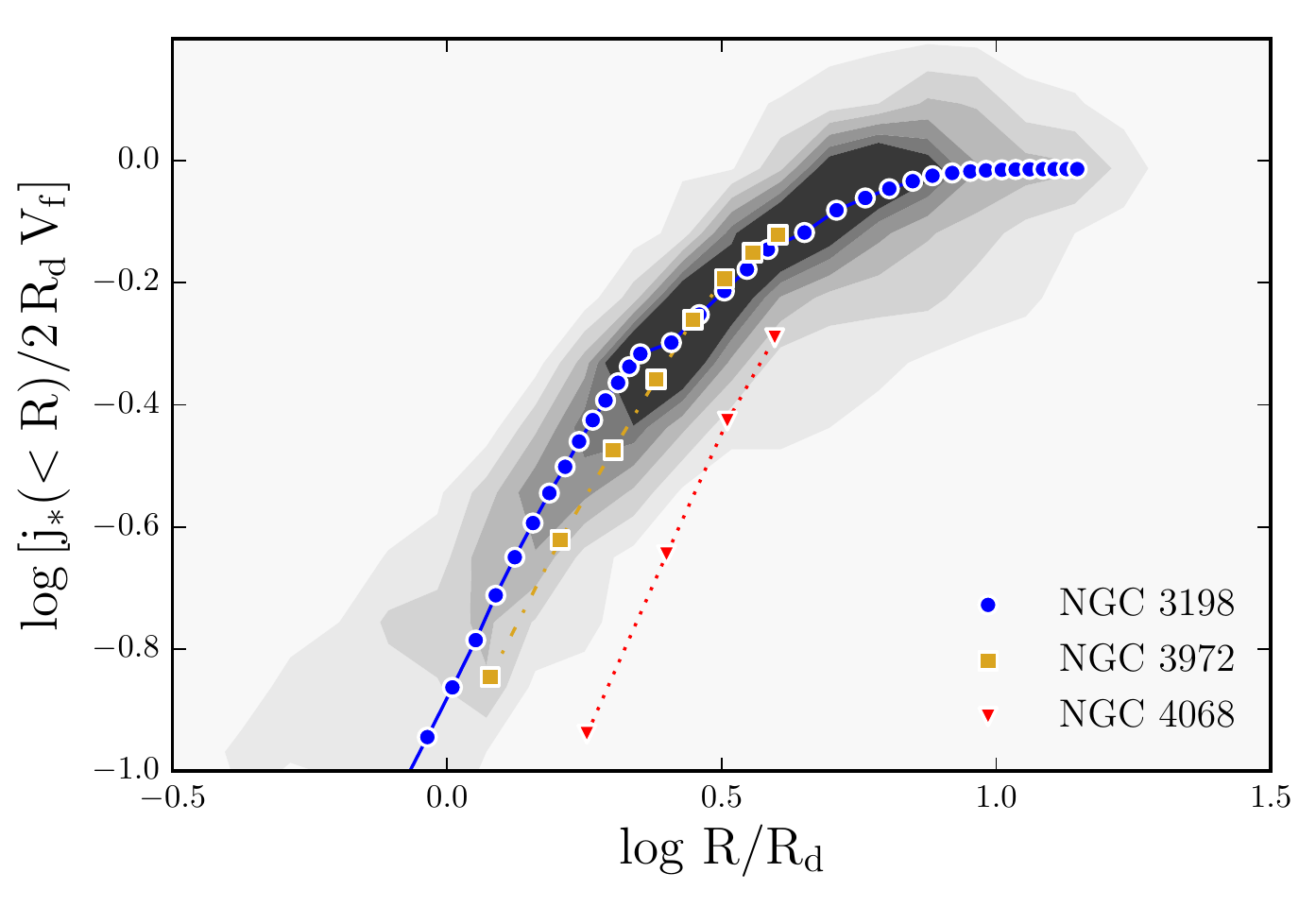}
\caption{Stellar specific angular momentum profiles for 175 disc galaxies in the
         SPARC sample. Grey contours represent the distribution of all the points in the
         profile of each galaxy. The radius is normalised to the disc scale length ($\Rd$ at
         3.6 $\mu$m) and $\jstar$ to that of a thin 
         exponential disc with the same $\Rd$ and with a constant rotation curve. 
         We also show the full profiles for three representative galaxies in our initial 
         sample: a galaxy with a fully converged $\jstar(<R)$ profile (blue circles), a galaxy 
         with a converging profile (yellow squares) and a galaxy with a non-converging profile 
         (red triangles). For our determination of the Fall relation, we excluded galaxies 
         with a non-converging profile.}
\label{fig:jstar_profiles}
\end{figure}

\begin{figure*}
\includegraphics[width=0.49\textwidth]{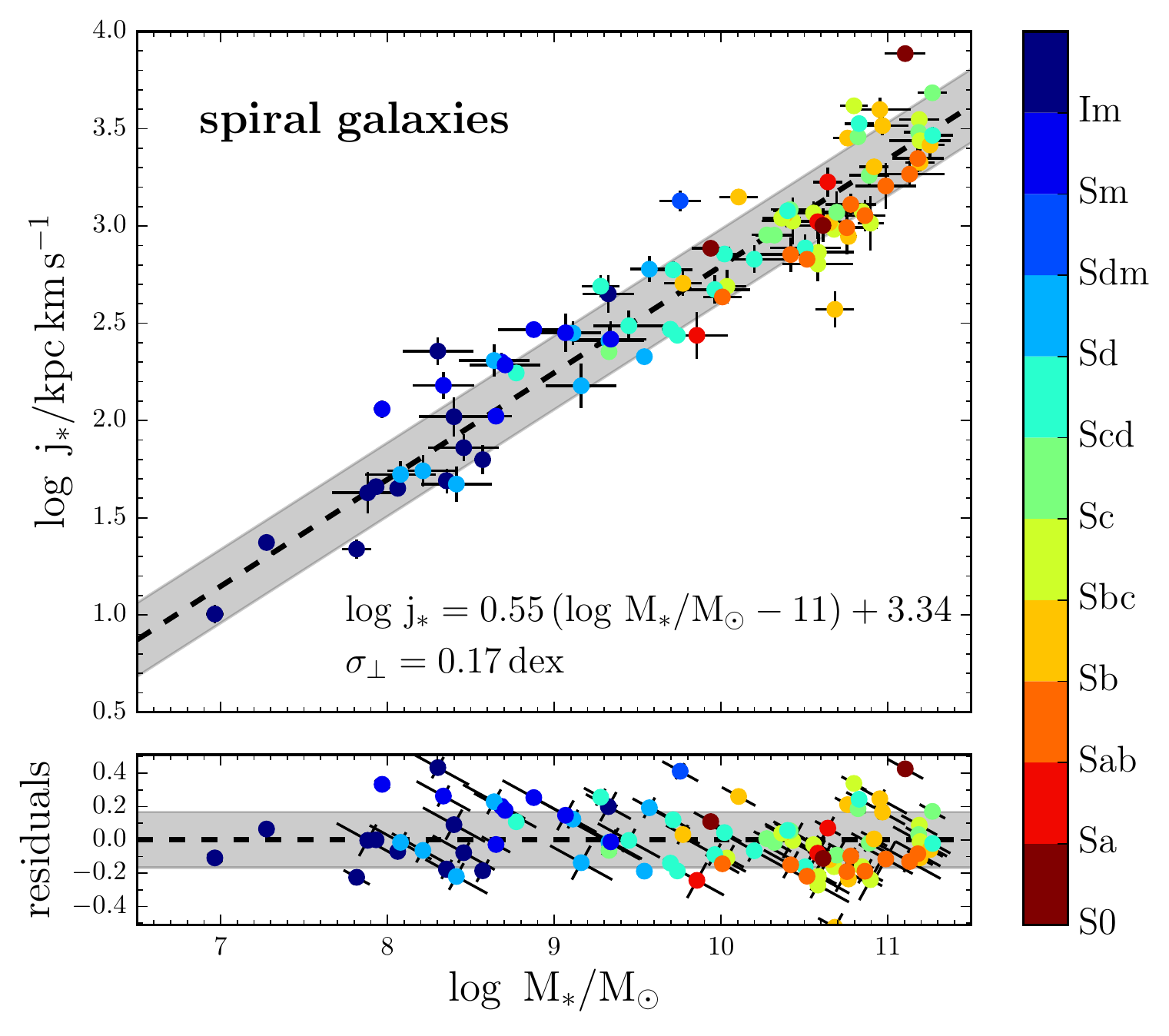}
\includegraphics[width=0.49\textwidth]{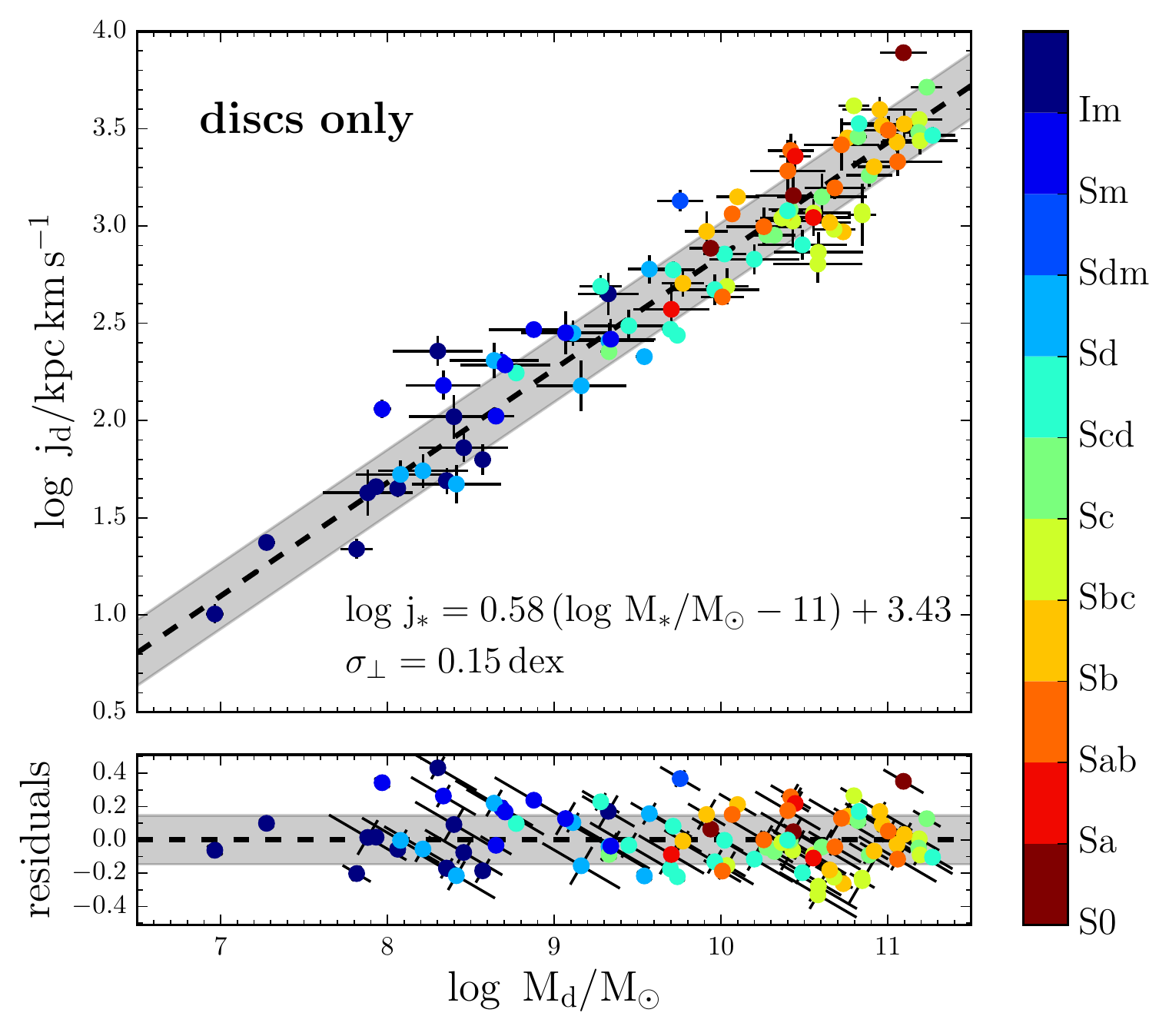}
\caption{\emph{Left-hand panel:} specific stellar angular momentum-stellar mass relation
		 (Fall relation) for a sample of 92 nearby disc galaxy. Each galaxy is represented by
         a circle coloured by Hubble type. 
         The black dashed line is the best-fitting linear model and
         the grey band is the $1\sigma$ orthogonal intrinsic scatter. The bottom panel shows
         the orthogonal residuals around the linear model.
         \emph{Right-hand panel:} same as the left-hand panel, but for the discs only (i.e.
         after removing the contribution from the bulges).
        }
\label{fig:fall}
\end{figure*}

\subsection{Specific angular momentum profiles and sample selection} \label{sec:jstar_profiles}

We use Eq.~\ref{eq:jstar} to compute the specific angular momentum as a function of radius.
We plot the resulting $\jstar(<R)$ profiles for all the galaxies in the SPARC sample in
Figure \ref{fig:jstar_profiles}. We find that for most galaxies $\jstar(<R)$ rises steeply
in the inner parts and then flattens at about $\sim 5\Rd$. The profiles typically flatten
close to the value of the specific angular momentum of a thin exponential disc (with scale
length $\Rd$) with a constant rotation curve $V_{\rm f}$, i.e. $j_{\ast,\rm exp} = 
2\Rd V_{\rm f}$.
As expected, most of the specific angular momentum of a spiral galaxy resides further out its
optical half-light radius and radially extended rotation curves, such as those provided by
\hi\ observations, are \emph{crucial} to properly measure it. In fact, all the galaxies with 
extended rotation curves have fully-converged $\jstar$ profiles, while only lower limits on
$\jstar$ can be determined for some galaxies for which the rotation curve at large radii is 
not known.

The value $j_{\ast,\rm exp} = 2\Rd V_{\rm f}$ has been used by several authors as an estimate
of the specific angular momentum for disc galaxies \citep[][\citetalias{RF12}]{Fall1983}.
However, the actual $j_\ast(<R)$ of spirals flattens at about $j_{\ast,\rm exp}$ with a 
significant scatter ($\sim 0.12$ dex at $\sim 5\Rd$), hence we caution against using simple 
estimates of the specific angular momentum.

To determine robustly the Fall relation for local spirals, we use only galaxies with an 
accurate measurement of $\jstar$, i.e. only those with a converging $\jstar(<R)$ profile.
Thus, being $R_0,\dots,R_N$ the radii at which the $\jstar$ profiles are sampled, we select
galaxies satisfying the following criteria:

\begin{equation} \label{eq:criteria}
\frac{\jstar(<R_N) - \jstar(<R_{N-1})}{\jstar(<R_N)} < 0.1 \quad \& \quad
\left.\frac{\de\log\jstar(<R)}{\de\log R}\right|_{R_N} < \frac{1}{2},
\end{equation}

\noindent i.e. that the last two points of the $\jstar$ profile differ by less than $10\%$ and
that the logarithmic slope of the $\jstar$ profile in the outermost point is less than $1/2$. 

To further illustrate our selection, we highlight in Fig.~\ref{fig:jstar_profiles} the individual
$\jstar$ profiles of three galaxies in the SPARC sample, representative of different converging
properties. NGC 3198 has a perfectly converged $\jstar$ profile (blue circles). In our sample, 28 
galaxies, covering the entire SPARC mass range, show a similar feature: these are the galaxies for
which $\jstar$ is best measured.
NGC 3972 (yellow squares) is the archetype of the galaxies that merely comply to the criteria of
Eq.~\ref{eq:criteria}; these galaxies have a converging $\jstar$ profile and 
the difference between the last two points of the profile is smaller than $10\%$.
NGC 4068 (red triangles) has a steeply rising $\jstar$ profile that does not meet 
Eq.~\ref{eq:criteria} criteria; 34 similar galaxies are excluded from the calibration of the 
local Fall relation since their total $\jstar$ might be severely underestimated.

We finally excluded galaxies with inclination angles below 30$^\circ$, as their rotation
velocity is very uncertain, and we are left with a sample of 92 galaxies with masses 
$7 \lesssim \log\Mstar/\Msun \lesssim 11.5$. For those, we estimate the uncertainty in the
stellar mass following \citet[][see their Section 2.3]{Lelli+2016} and the error on $\jstar$
as

\begin{equation}
\delta_{\jstar} = \Rd\sqrt{\frac{1}{N}\sum_i^N\delta_{v_i}^2 + 
      \left(\frac{\Vf}{\tan\,i}\delta_i\right)^2 +
      \left(\Vf\frac{\delta_D}{D}\right)^2 },
\end{equation}

\noindent where $\Vf$ is the velocity in the flat part of the rotation curve \citep[see][]
{Lelli+2016}, $i$ is the inclination and $\delta_i$ its uncertainty, $D$ is the distance and
$\delta_D$ its uncertainty and $\delta_{v_i}$ is
the uncertainty at each point in the rotation curve. The error on distance often dominates 
the error budget. Of the 92 galaxies selected, 49 (53$\%$) have relatively uncertain
distances estimated with the Hubble flow (with relative errors of $10-30\%$), while 43
(47$\%$) have distances known within better than $10\%$ (mostly from red giant branch tip).

\subsection{The $\jstar-\Mstar$ relation for galaxies and their discs} \label{sec:jstar_mstar}

The left-hand panel of Figure~\ref{fig:fall} shows our determination of the specific
angular momentum-mass relation for nearby disc galaxies over $\sim 5$ dex in stellar mass. 
We fit a linear relation (in logarithm) to the data points allowing for an \emph{orthogonal} 
intrinsic\footnote{We subtract the contribution to the  total scatter from measurement 
uncertainties.} scatter. We assume uninformative priors for the three parameters 
(slope, normalisation and scatter) and explore the
posterior distribution with a Monte Carlo Markov Chain (MCMC) method \citep[using the 
\texttt{python} implementation by][]{Foreman-Mackey2013}.
With a model that follows

\begin{equation}
\log\jstar = \alpha\,[\log(\Mstar/\Msun) - 11] + \beta,
\end{equation}

\noindent we find a best-fitting slope $\alpha = 0.55 \pm 0.02$, a normalisation $\beta = 3.34
\pm 0.03$ and an orthogonal intrinsic scatter $\sigma_\perp = 0.17 \pm 0.01$ dex. We repeated
this exercise i) varying the thresholds in Eq.~\ref{eq:criteria} and ii) considering only
the 20 galaxies with converged $\jstar$ profiles that have distances known better than $10\%$, 
and found no significant difference in the best-fit relation. 
In these estimates we have assumed the uncertainties in $\Mstar$ and $\jstar$ to be
uncorrelated; however, this is likely not the case since both $\delta_{\Mstar}$ and
$\delta_{\jstar}$ are often dominated by distance errors. 
Hence, we recomputed again the distributions of the model
parameters in the extreme case of fully correlated uncertainties (correlation coefficient unity):
we find no significant difference in neither the slope nor the normalisation, but we find
a slightly larger orthogonal intrinsic scatter $\sigma_\perp = 0.179\pm0.014$ dex.

The best-fitting values are consistent, albeit having a smaller intrinsic scatter, 
with previous estimates of the Fall relation for high-mass spirals. We also confirm that
the residuals correlate with galaxy morphology: earlier galaxy types are found 
systematically below the relation and viceversa for later types \citep[\citetalias{RF12};]
[]{Cortese+2016}.
While significantly improving the determination of the relation at
high masses\footnote{\citetalias{RF12} used the simple $j_{\ast, \rm exp}$ estimator, 
\cite{Cortese+2016} computed $\jstar$ only within the optical effective radius and
\cite{OG14} had only 16 objects.}, \emph{we have robustly measured that
the Fall relation extends to dwarf galaxies as a single, unbroken power-law}.
This is a crucial observational result that challenges many 
state-of-the-art galaxy formation models, which predict a flattening of the relation at
low masses \citep{Stevens+2016,Obreja+2016,Mitchell+2018}.

We note that two previous works looked at the \emph{baryonic} version of the specific
angular momentum-mass law for dwarf irregulars and found them to be offset towards larger
$j_{\rm baryon}$ with respect to the relation for massive spirals
\citep{Butler+2017,ChowdhuryChengular2017}. 
Our rotation curves (from the SPARC sample) have been specifically selected to be of the 
highest possible quality and hence to better trace the axisymmetric gravitational potential.
We are planning to use this dataset to investigate the baryonic relation in a forthcoming
paper. \cite{Butler+2017} also showed a $\jstar-\Mstar$ relation down to low-mass galaxies
which is offset towards larger $\jstar$ and much more scattered than ours. This
is most likely due to i) the different quality of their rotation curves \citep[see e.g.]
[their Fig.\ 24]{Iorio+2017} and ii) to their use of fitting functions to extrapolate the
total $\jstar$ as opposed to our check of its convergence on each individual galaxy.
Moreover, we have checked our results against a different determination of the circular 
velocities for a sample of 17 dwarf irregulars by \cite{Iorio+2017}, who used the
state-of-the-art 3D software {\sc 3d-barolo} \citep{Barolo}. 
We found that the $\jstar$ profiles of the galaxies in common with SPARC are consistent.

\begin{table}
\caption{Best-fit parameters, and their $1\sigma$ errors, of the function
		 $\log j = \alpha\,(\log M - 11) + \beta$ fitted to the data in 
         Fig.~\ref{fig:fall} for galaxies and their discs. $\sigma_\perp$ is the orthogonal
         intrinsic scatter.}
\label{tab:fits}
\small
\begin{tabular}{lccc}
 & $\alpha$ & $\beta$ & $\sigma_\perp$ \vspace{.1cm} \\
\hline
{\bf Spiral galaxies} & $0.55 \pm 0.02$ & $3.34 \pm 0.03 $ & $0.17 \pm 0.01$ \\
{\bf Discs only} & $0.585 \pm 0.020$ & $3.43 \pm 0.03$ & $0.15 \pm 0.01$ \\
\end{tabular}
\end{table}

The Fall relation in the left-hand panel of Fig.\ref{fig:fall} is amongst the tightest known
scaling laws for spiral galaxies\footnote{For comparison the baryonic Tully-Fisher relation
has an intrinsic orthogonal scatter of about $\sim 0.07-0.05$ dex \citep{Ponomareva+2018}.}.
However, the relation gets even tighter if one considers just the disc component of spiral 
galaxies, i.e. by removing the bulge contribution to the light profile. 
The right-hand panel of Fig.~\ref{fig:fall} shows the disc specific angular momentum-mass 
relation ($\jd-\Md$) and Table~\ref{tab:fits} summarises the results for both the 
$\jstar-\Mstar$ and $\jd-\Md$ relation. 
We find that the relation for discs has the following important differences 
with respect to the one for whole stellar body:
\begin{itemize}
\item S0-Sb galaxies with $\log\Mstar/\Msun\gtrsim 9.5$ scatter around null residuals in the
      $\jd-\Md$ relation. This significantly alleviates the trend of the residuals with galaxy 
      morphology, present in the $\jstar-\Mstar$ relation;
\item the scatter of the $\jd-\Md$ relation, $\sigma_\perp=0.15\pm 0.01$ dex,
      is slightly smaller and its slope, $\alpha=0.585\pm 0.020$, is 
      slightly larger than that of the $\jstar-\Mstar$ relation.
\end{itemize}
These two points are particularly important because they suggest that the $\jd-\Md$ is
more fundamental than the $\jstar-\Mstar$ relation. In particular, the best-fit slope of 
the relation for discs is closer to that of dark matter haloes, $\jh\propto \Mh^{2/3}$, 
possibly indicating a simpler link to dark haloes of discs compared to bulges 
\citep[e.g.][]{MMW98}. In the next Section we discuss
some implications of the $\jd-\Md$ relation for galaxy formation models.

\begin{figure}
\includegraphics[width=0.49\textwidth]{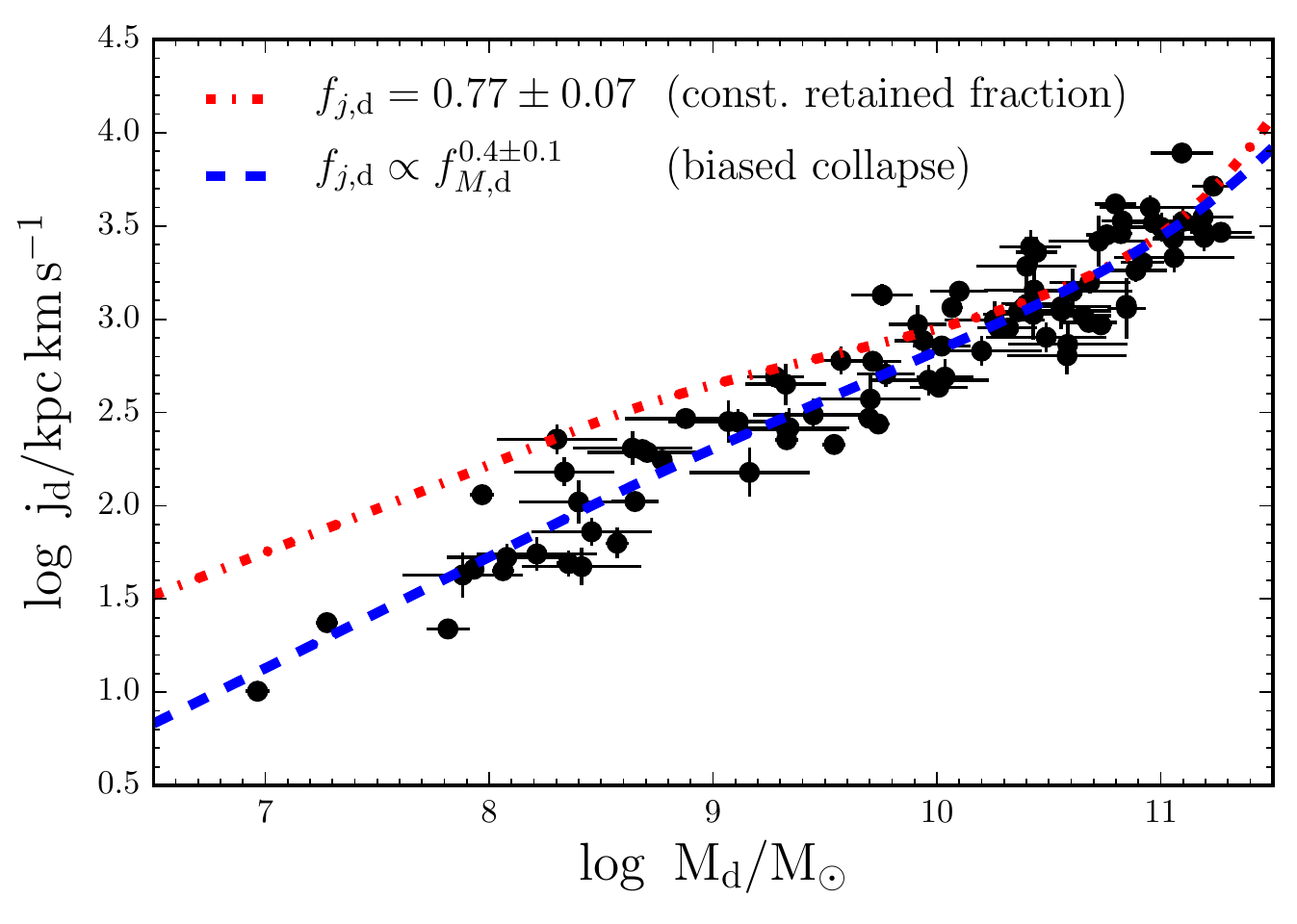}
\caption{Predicted distribution in the $\jd-\Md$ plane of a model with
         a constant retained fraction of angular momentum $f_j$ (red dot-dashed line) and 
         for a \emph{biased collapse} model (blue dashed line) compared to the data as in 
         Fig.~\ref{fig:fall}.
         To compute the two models we have used the stellar-to-halo mass relation from 
         \citet{RodriguezPuebla+2015}.}
\label{fig:interp}
\end{figure}

\section{Physical models} \label{sec:models}

In a $\Lambda$CDM cosmology, dark matter haloes acquire angular momentum from tidal torques
such that their specific angular momentum $\jh \propto \lambda\Mh^{2/3}$, being $\Mh$ their
(virial) mass and $\lambda$ the so-called dimensionless spin parameter \citep[which is
independent from mass, see][]{Peebles69}.
It follows that discs, which form inside these haloes out of gas that has initially
the same initial angular momentum as the dark matter, will have specific angular momentum

\begin{equation} \label{eq:jd_Md_fj}
\jd \propto \lambda \fjd \fmd^{-2/3} \Md^{2/3},
\end{equation}

\noindent where we have defined the stellar disc's ``global star-formation efficiency'' 
$\fmd \equiv \Md/\Mh$ and the stellar disc's ``retained fraction of angular momentum'' 
$\fjd \equiv \jd/\jh$ \citep[e.g.][]{Posti+2017}. 
In what follows we will use the recent estimate of the
stellar-to-halo mass relation for late-type galaxies by \cite{RodriguezPuebla+2015} as our
fiducial $\fmd$ \citep[see][for other choices]{Posti+2017}.

We first consider the case in which all spirals retain the same fraction of the halo's
angular momentum \citep[e.g.][]{MMW98}. This case is of particular interest since several 
state-of-the-art semi-analytic galaxy formation models are based on the assumption of a 
constant $\fjd$ for all masses 
\citep[see e.g.][for a comparison of many of these]{Knebe+2015}.
In this case, Eq.~\ref{eq:jd_Md_fj} reduces to $\jd \propto \fmd^{-2/3}\Md^{2/3}=\Mh^{2/3}$,
which we show in Figure~\ref{fig:interp} in comparison with the data (red dot-dashed line).
We fitted this model for
the value of the constant retained fraction and we found it significantly smaller than unity, 
$\fjd=0.77 \pm 0.7$ \citep[similar to previous findings, e.g.][\citetalias{RF12}]
{DuttonvandenBosch2012}. While the model provides a decent fit to the data for high-mass
discs ($\log\Mstar/\Msun\sim 10.5$), indicating that about $70-80\%$ of the halo's angular
momentum is incorporated in these galaxies, the model also clearly over-predicts $\jd$ in 
low-mass galaxies ($\log\Mstar/\Msun\lesssim 9$), which instead require a much smaller $\fjd$ 
\citep[see also Figure 1 in][]{Posti+2017}.

Some semi-analytic galaxy formation models \citep{Knebe+2015,Stevens+2016} as well as some
others based on numerical hydrodynamical simulations 
\citep{Genel+2015,Obreja+2016,Mitchell+2018} do show similar predictions of a flattening of
the Fall relation at $\Mstar\sim 10^9\Msun$, which are  inconsistent with our accurate 
determination of the $\jstar-\Mstar$ law. This is, in fact, related to the retained fraction
of angular momentum  $\fjd$ which they either assume or find to be relatively constant as a
function of galaxy mass. 
In other models, instead, strong stellar feedback is crucial to make low-mass discs have a 
lower $\fjd$ -- which also decreases to smaller masses -- and are in fact able to predict 
qualitatively well the Fall relation in the full mass range that we probe observationally 
\citep{El-Badry+2018}.

The last physical models that we consider in this discussion are inside-out cooling models in
which star-formation proceeds from angular momentum-poor to angular momentum-rich gas:
the so-called \emph{biased collapse} models \citep[\citetalias{RF12}]{DuttonvandenBosch2012,
Kassin+2012}.
Following \cite{Posti+2017}, if for the gas, as for the dark matter, the angular momentum
is distributed as $j(<R) \propto M(<R)^s$, with $s$ being a constant slope \citep[see][]
{Bullock+2001}, then $\fjd\propto\fmd^s$ \citep{vandenBosch1998,NavarroSteinmetz2000}. 
Plugged into Eq.~\ref{eq:jd_Md_fj}, this yields $\jd\propto\fdisc^{s-2/3}\Md^{2/3}$.
We fit this prediction for the best value of $s$, which we find to be $s=0.4\pm 0.1$, and we
show how that compares to the observed Fall relation in Fig.~\ref{fig:interp} (blue dashed 
line). This model
provides a remarkable fit to the observations over the entire mass range and it is preferred
to the constant-$\fjd$ one (according to the Bayesian Information Criterion). The slope $s$
in this model is related to the canonical angular momentum distribution $\de M/\de j\propto 
j^q$ where $s=1/(1+q)$ and, perhaps unsurprisingly, its best-fitting value is compatible with 
that expected for pure discs \citep[$q\simeq 1$ in the low-$j$ regime,][]{vdBBS2001}.

Finally we note that the scatter that we measure for the Fall relation ($\sigma_\perp=0.15$ 
dex) is significantly smaller than what one might expect from Eq.~\eqref{eq:jd_Md_fj},
given that the scatter in the $\lambda$ distribution is $\sim 0.25$ dex and that in the
stellar-to-halo mass relation is $\sim 0.15$ dex. This suggests that i) the scatters
of the $\lambda-\Mh$, $\fjd-\Mh$ and $\Md-\Mh$ relations are correlated such that when
combined in Eq.~\eqref{eq:jd_Md_fj} they yield the observed scatter and/or ii) the $\lambda$
distribution of haloes hosting spiral galaxies is intrinsically narrower than that of the
full halo population \citep[even if the latter can not alone explain the measured specific
angular momenta of massive spirals and ellipticals, see][]{Posti+2017}.

\section{Summary \& Conclusions} \label{sec:concl}

In this Letter we have studied the relation between specific angular momentum and mass (Fall
relation) of nearby disc galaxies spanning an unprecedented range in stellar mass
($7 \lesssim \log\Mstar/\Msun \lesssim 11.5$). We have used {\it Spitzer} 3.6 $\mu$m photometry
and \hi\ rotation curves, compiled in the SPARC sample, to trace respectively the stellar
mass surface density and rotation velocity profiles. We determine specific angular momentum
profiles for all galaxies and use only those with converging profiles in the determination
of the scaling law, since they guarantee an accurate measurement of the total galactic
angular momentum. We find that:
\begin{itemize}
\item[(i)] the Fall relation, $\jstar-\Mstar$ is remarkably well represented
           by a single power-law, with slope $\alpha=0.55\pm 0.02$ and scatter 
           $\sigma_\perp=0.17\pm 0.01$ dex, from massive to dwarf spiral galaxies;
\item[(ii)] the disc-only relation $\jd-\Md$ has a slightly steeper slope 
            $\alpha=0.585\pm 0.020$ and a slightly smaller scatter $\sigma_\perp=0.15\pm 0.01$ 
            dex;
\item[(iii)] the observed Fall relation is a powerful benchmark for galaxy formation
             scenarios and poses a challenge to some of the current models, which predict a 
             change of slope at low masses.
\end{itemize}

Being a scaling law that tightly relates two fundamental and independent quantities subject
to physical conservation laws, the Fall relation stands out as one of the most (if not
\emph{the} most fundamental scaling relation for disc galaxies. From dwarf irregulars to
massive spirals, from early-types (S0) to late-type spirals (Sd-Sm), all disc galaxies in
the local Universe appear to lie on a single power-law relation, which may already be in place 
in the early Universe \citep{Burkert+2016,Harrison+2017}.

We also discussed how these remarkable observations can be used to constrain galaxy formation 
models. In particular, the Fall relation uniquely constrains how much of the angular momentum
initially present in the baryons ends up being encapsulated in the stellar body of a galaxy.
Models assuming this to be a constant with galaxy mass will inevitably fail at
reproducing the observations. 

Among those considered, an inside-out cooling model (biased collapse) works better in 
reproducing the observed law. This scenario also points to some angular momentum
redistribution taking place during star formation, possibly due to feedback, or to
significant differences between the angular momenta of dark matter and baryons at start. 
Understanding the precise nature of these processes and in general giving account of the
observed straight and tight $j_\star-M_\star$ relation across almost 5 orders of magnitude
in stellar mass is a key challenge of modern theoretical astrophysics.

\appendix
\section{Asymmetric drift correction} \label{sec:ad}

The circular velocity $\Vc$ is equal to the sum in quadrature of the stellar rotation
velocity $V_{\rm \ast,rot}$ and the asymmetric drift velocity $V_{\rm AD}$, i.e.
$\Vc^2 = V_{\rm \ast,rot}^2+V_{\rm AD}^2.$
We use the (inclination-corrected) \hi\ observed rotation curve to trace $\Vc$
at each radius. Then, following the findings of the DiskMass Survey on a sample
of 30 well-studied face-on nearby spirals \citep{Martinsson+2013}, we assume the
vertical stellar velocity dispersion to vary exponentially with radius $\sigma_z = \sigma_{0,z}
\exp(-R/2\Rd)$, where $\Rd$ is the disc scale length measured at 3.6 $\mu$m. The
normalisation $\sigma_{0,z}$ is found to be a function of the galaxy's central surface 
brightness $\mu_0$ at 3.6 $\mu$m: $\sigma_{z,0}\simeq 20$ km/s for $\mu_0\lesssim 20$ 
mag/arcsec$^2$ and $\sigma_{z,0}\simeq 70$ km/s for $\mu_0 \sim 16$ mag/arcsec$^2$ 
\citep[see][]{Martinsson+2013}.

By further assuming that the disc scale height is constant with radius and that the system
is isotropic, $\sigma_R=\sigma_z$, we can write the asymmetric drift velocity as
\citep[e.g.][\S 4.8.2]{BT08}
\begin{equation} \label{eq:ad}
V_{\rm AD}^2 = \sigma_{0,z}^2 \frac{3R}{2\Rd} e^{-R/2\Rd}.
\end{equation}
In this Letter we use Eq.~\ref{eq:ad} to correct for asymmetric drift. In general,
this introduces just a small correction, of less than $5\%$, to the estimate of the specific
angular momentum. We have also considered i) the anisotropic case $\sigma_z^2=\sigma_R^2/2$ and
ii) the extreme case of a uniform vertical dispersion $\sigma_z=\sigma_{0,z}$ throughout the 
galaxy and found in both cases small differences (less than $10\%$) in the derived specific 
angular momenta.

\end{document}